\documentclass[12pt]{article}
 \usepackage{amsmath,amsxtra}
 \usepackage{amssymb}
 \usepackage{bm}
 \usepackage{graphicx}
 \usepackage{graphics}
 \usepackage{epsfig,subfigure}
 \usepackage{cite}

\title{Lie group analysis for multi-scale plasma dynamics}

\author{Vladimir F.~Kovalev \\
 \small{\textit{Keldysh Institute of Applied Mathematics, Miusskaya Pl.,4 }}\\
 \small{\textit{ Moscow, 125047, Russia, e-mail: vkovalev@imamod.ru}}}

\date{}

\begin{document}
\maketitle

\begin{abstract}
An application of approximate transformation groups to study dynamics of a system with distinct
time scales is discussed. The utilization of the Krylov-Bogoliubov-Mitropolsky method of
averaging to find solutions of the Lie equations is considered. Physical illustrations from the
plasma kinetic theory demonstrate the potentialities of the suggested approach. Several examples
of invariant solutions for the system  of the Vlasov-Maxwell equations for the two-component
(electron-ion) plasma are presented.
\end{abstract}

\section{Introduction}

In analyzing various physical systems we frequently deal with a situation when a complicated
dynamics of different systems appears as a superposition of ``fast'' and ``slow'' motions with
incommensurable characteristic scales, for example, slow evolution of ``background'' system
characteristics accompanied by fast oscillations in the vicinity of a background state. This type
of behavior seems typical for various linear and nonlinear problems (numerous examples are found
in \cite{Nayfeh_bk_73,ZhurKlymov_bk_88}), e.g. for celestial mechanics in studies of a motion of
planets, for mechanics when treating oscillatory regimes of systems with slowly varying
parameters, for various nonlinear problems of multi-component plasma.

The availability of different scales (though the origin of these scales depends upon the
particular system of interest) allows to simplify the analysis of the complicated dynamics by
treating ``fast'' and ``slow'' motions separately. These ideas underlie an essence of asymptotic
analytical approaches, the method of averaging \cite{Krylov_bk_37,KBM_bk_74}, the method of
multiple scales \cite{Nayfeh_bk_73}, and other asymptotic methods (see, e.g.,
\cite{Nayfeh_bk_73,ZhurKlymov_bk_88}).

As to relation of modern group analysis to nonlinear dynamics here we will point to an
interpenetration of ideas from both fields: on one hand the use of the Lie group theory in
asymptotic methods for integration of nonlinear differential equations gives (in combination with
the Hausdorff formula) the theoretical basis for the method of averaging \cite{KBM_bk_74} and
provides a regular procedure for calculating the asymptotic series in this method
\cite{Hori_pasj_66,ZhurKlymov_bk_88,Lopatin_bk_88}. On the other hand an introduction of
multiple-scales approach to modern group analysis \cite{BGI_pr_91} enhances the potentiality of
approximate transformation groups \cite{BGI_matsb_89}.

The present paper uses the Krylov-Bogoliubov-Mitropolsky method (KBM-method) of averaging in
group analysis of the system of equations that describes the evolution of plasma particles in
multi-component plasma. The linearity of the group determining equations plays the decisive role
in separating fast and slow terms in coordinates of a group generator and in successive use of
the KBM-method for constructing the  asymptotic solutions of the Lie group equations.

The paper is organized as follows: in Section~\ref{sec2} we introduce the basic equations for
evolution of a two component plasma and introduce small parameters which give rise to different
time and spatial scales for distinct plasma components. In the next Section~\ref{sec3} we
describe the solution of determining equations, which define approximate Lie point symmetry
group. The use of KBM-method in finding solutions of the Lie group equations constitutes the
basis of the Section~\ref{sec4}. Several examples of an application of the suggested approach are
presented in Section~\ref{sec5} and Section~\ref{sec6}. In the Conclusion we discuss the results
obtained and the future application of the KBM-method in modern group analysis.

\section{Basic Equations: Electron-ion Plasma \label{sec2}}

We start with kinetic equations for distribution functions, $f^e$ and $f^i$,
 \begin{equation}
  \begin{aligned}
  & \partial_t f^{e}  + v^{e} \partial_x f^{e}
  - (e/m) E \partial_{v^{e}} f^{e}  = 0\,, \quad
  \partial_t f^{i}  + v^{i} \partial_x f^{i}
  + (Ze/M) E \partial_{v^{i}} f^{i}  = 0\,,
  \end{aligned}
  \label{kineq}
 \end{equation}
for both species of two-component plasma consisting of electrons and ions with mass $m$ and $M$ and charges $e^e=-e$ and
$e^i=Ze$, where $Z$ is a charge number, and equations for a self-consistent electric field $E$
 \begin{equation}
  \begin{aligned} &
 \partial_x E=4\pi \rho \,, \quad  \partial_t E = - 4\pi j \,, \quad  \partial_t \rho + \partial_x j = 0 \,.
  \end{aligned}
  \label{elfield}
 \end{equation}
Here charge $\rho$ and current $j$ densities are related to moments of the distribution functions
via nonlocal material relations:
 \begin{equation}
  \begin{aligned} &
 \rho=  e \left[ Z \int\textrm{d} v^{i} f^{i} -
 \int\textrm{d} v^{e} f^{e} \right] \, , \quad
 j = e \left[ Z \int\textrm{d} v^{i} v^{i} f^{i} -
 \int\textrm{d} v^{e} v^{e} f^{e} \right] \, .
  \end{aligned}
  \label{material}
 \end{equation}
Equations (\ref{kineq})-(\ref{material}) are known as a system of the Vlasov-Maxwell equations
for a collisionless plasma. We are interested in the solution of the Cauchy problem for kinetic
equations (\ref{kineq}) with the initial conditions
  \begin{equation}
  f^{e}|_{t=0}=f^{e}_0(x,v^{e}) \,, \quad  f^{i}|_{t=0}=f^{i}_0(x,v^{i}) \,,
  \label{initialcond}
 \end{equation}
which depend on a particular physical problem. In what follows we consider an evolution of
localized plasma bunches and assume sufficiently smooth (e.g., Maxwellian) initial distribution
functions with electron $T_e$ and ion $T_i$ temperatures and initial densities of electrons
$n^{e}(x) = \int\textrm{d} v^{e} f^{e}_{0}$ and ions $n^{i}(x) = \int\textrm{d} v^{i} f^{i}_{0}$
with the characteristic scale $L$. Below we consider a typical situation when $L$ is much greater
than the Debye radius of electrons $r_{De}=\sqrt{T_e/(4 \pi n^e_0 e^2)}$. The difference in mass
of plasma particles specifies two different time scales, namely dimensionless time $\omega_{Le}
t$ for ``fast'' electron motions and $\tau=\mu t$ for ``slow'' motions, $\mu=\sqrt{Zm/M} \ll 1$.
It is natural that electrons are involved in both fast and slow motions, hence the electron
distribution function depends on both $t$ and $\tau$. On the contrary, we consider ions not
involved in fast motions. It means that the ion distribution function does not depend upon fast
time $t$, but only upon slow time $\tau$. It also means that averaging upon fast time eliminates
the fast component $\tilde E$ of the electric field $E={\tilde E} + {\bar E}$ in the ion kinetic
equation that contains only the slow electric field $\bar E$. Then introducing dimensionless
variables, electron velocity $u=v^{e}/V_{Te}$, $V_{Te}=\sqrt{T_e/m}$, dimensionless ion velocity
$w=v^{i}/c_{s}$, $c_s =\sqrt{ZT_e/M}$, dimensionless electric field $p=\varepsilon (eEL/T_e)$\,,
$\varepsilon =r_{De}/L \ll 1$ and dimensionless distribution functions $f^e=(n^{e}_{0}/V_{Te}) g
$, $f^i=(n^{e}_{0}/(Zc_s)) f$, $n^{e(i)}_{0}=n^{e(i)}(0) $ we come to the following system of
basic equations in dimensionless variables
 \begin{align} &
 \partial_t g + \mu \partial_\tau g  + \varepsilon u \partial_x g - p \partial_{u} g  = 0\,, \quad
  \partial_\tau f + \varepsilon w \partial_x f + \bar{p} \partial_{w} f  = 0 \,,
  \label{basiceq_local}
 \\ &
 \varepsilon \partial_x p = \int\textrm{d} w f - \int\textrm{d} u g \,, \quad
  \partial_t p + \mu \partial_\tau  p = - \mu \int\textrm{d} w w f + \int\textrm{d} u u g  \,.
 \label{basiceq_nonlocal}
 \end{align}
From the group analysis point of view this system of equations should be supplemented by the four
additional equalities,
 \begin{equation}
 \partial_w g = 0 \,, \quad  \partial_u f = 0 \,, \quad \partial_u p = 0 \,, \quad  \partial_w p = 0 \,,
 \label{basiceq2}
 \end{equation}
which are evident from the physical point of view.

\section{Lie symmetry group \label{sec3}}

The Lie point symmetry group admitted by the system (\ref{basiceq_local}) and
(\ref{basiceq_nonlocal}) is defined by a symmetry group generator
 \begin{equation}
  X = \xi^1 \partial_{t} +
     \xi^2 \partial_{x} +
     \xi^3 \partial_{u} +
     \xi^4 \partial_{w} +
     \xi^5 \partial_{\tau} +
     \eta^1 \partial_g +
     \eta^2 \partial_p +
     \eta^3 \partial_f  \,.
 \label{gen1}
 \end{equation}
In the canonical form this generator is written as
 \begin{equation}
  Y = \kappa^1 \partial_{g} +
     \kappa^2 \partial_{p} +
     \kappa^3 \partial_{f} \,,
 \label{gen1_canonical}
 \end{equation}
\[
 \begin{aligned} &
 {\kappa}^{1} = {\eta}^{1} -{\xi}^{1} \partial_t g -{\xi}^{2} \partial_x g
                -{\xi}^{3} \partial_u g -{\xi}^{5} \partial_\tau g \,, \\ &
 {\kappa}^{2} = {\eta}^{2} -{\xi}^{1} \partial_t p -{\xi}^{2} \partial_x p
                -{\xi}^{5} \partial_\tau p \,, \\ &
 {\kappa}^{3} = {\eta}^{3} -{\xi}^{2} \partial_x f
                -{\xi}^{4} \partial_w f -{\xi}^{5} \partial_\tau f \,.
 \end{aligned}
  \]
When applying group generator (\ref{gen1_canonical}) to (\ref{basiceq_local}),
(\ref{basiceq_nonlocal}) and (\ref{basiceq2}) we get a system of determining equations for
coordinates $\xi^i$, $\eta^i$ of the generator (\ref{gen1}),
 \begin{equation}
 \begin{aligned} &
  D_t {\kappa}^1 + \mu D_\tau {\kappa}^1 + \varepsilon u D_x {\kappa}^1 - p D_u {\kappa}^1 - {\kappa}^2
  \partial_u g = 0 \,, \\ &
  D_\tau {\kappa}^3 + \varepsilon w D_x {\kappa}^3 + \bar p D_w {\kappa}^3 - {\bar{\kappa}}^2
  \partial_w f = 0 \,, \\ &
  D_w {\kappa}^1 = 0\,, \quad D_u {\kappa}^3 =0 \,, \quad D_u {\kappa}^2 =0 \,, \quad D_w {\kappa}^2 =0 \,,
  \end{aligned}
 \label{detereq_gen_local}
\end{equation}
 \begin{equation}
 \begin{aligned} &
  \varepsilon D_x {\kappa}^2 - \int\textrm{d} w {\kappa}^3 + \int\textrm{d} u {\kappa}^1  = 0 \,, \\ &
  D_t {\kappa}^2 + \mu D_\tau {\kappa}^2 + \mu \int\textrm{d} w w {\kappa}^3
  - \int\textrm{d} u u {\kappa}^1 = 0\,,
  \end{aligned}
 \label{detereq_gen_nonlocal}
\end{equation}
which should be solved in view of the basic equations (\ref{basiceq_local}),
(\ref{basiceq_nonlocal}), (\ref{basiceq2}) and all their differential consequences. Here $D_t$,
$D_\tau$, $D_x$, $D_u$ and $D_w$ are total differentiations with respect to the variable denoted
by lower index,
\begin{equation}
 \begin{aligned} &
  D_t = \partial_t + (\partial_t g)\, \partial_g + (\partial_t p)\, \partial_p \,, \\ &
  D_\tau = \partial_\tau + (\partial_\tau g)\, \partial_g + (\partial_\tau f)\, \partial_f
     + (\partial_\tau p)\, \partial_p \,, \\ &
  D_x = \partial_x + (\partial_x g)\, \partial_g + (\partial_x f)\, \partial_f
     + (\partial_x p)\, \partial_p \,, \\ &
  D_u = \partial_u + (\partial_u g)\, \partial_g \,, \quad
  D_w = \partial_w + (\partial_w f)\, \partial_f \,.
  \end{aligned}
 \label{differ}
\end{equation}
To find coordinates $\xi^i$, $\eta^i$ from the system of determining equations we use the
approach developed in \cite{kov-de-93} (see also \cite[Chapter 4]{GrIbrKovMel_book_2010}).
Following this technique we separate the determining equations for $\xi^i$ and $\eta^i$ into
local determining equations, (\ref{detereq_gen_local}), which arise from invariance of
(\ref{basiceq_local}), (\ref{basiceq2}), and nonlocal determining equations,
(\ref{detereq_gen_nonlocal}), which follow from invariance conditions for
(\ref{basiceq_nonlocal}). Solutions of local determining equations give the so-called
intermediate symmetry \cite{kov-de-93}. \par

Two distinct moments should be taken into account here: first, in view of multi-scale dynamics we
outline in coordinates $\xi^1$, $\xi^2$, $\xi^3$ and $\eta^2$, entering local determining
equations the fast terms denoted by variables with tilde and slow terms denoted by variables with
bar
 \begin{equation}
 \xi^i = {\tilde{\xi}}^i + {\bar{\xi}}^i \,, \quad i=1,2,3\,,  \qquad
 \eta^2 = \tilde{\eta}^2 + \bar{\eta}^2  \,.
 \label{fastandslow}
 \end{equation}
Due to the fact that both local and nonlocal determining equations are linear in $\xi$ and
$\eta$, we thus can separate terms of different characteristic scales. Then omitting trivial
tedious computations we rewrite fast
 \begin{equation}
 \begin{aligned} &
 {\tilde \eta}^2 +p \left( \partial_t {\tilde \xi}^1 + \mu \partial_{\tau} {\tilde \xi}^1
 - \partial_{u} {\tilde \xi}^3 \right)
 + \partial_t {\tilde \xi^3} + \mu \partial_{\tau}{\tilde \xi}^3 + \varepsilon u p \partial_{x}{\tilde \xi}^1
 + \varepsilon u \partial_{x}{\tilde \xi}^3 = 0 \,, \\ &
 \varepsilon {\tilde \xi}^3 - \partial_{t}{\tilde \xi}^2 - \mu \partial_{\tau}{\tilde \xi}^2
 + \varepsilon u \left( \partial_{t} {\tilde \xi}^1 + \mu \partial_{\tau}{\tilde \xi}^1
 - \partial_{x}{\tilde \xi}^2 \right)
 + \varepsilon^2 u^2 \partial_{x} {\tilde \xi}^1 = 0 \,,
  \end{aligned}
 \label{detereq1}
 \end{equation}
and slow local determining equations,
 \begin{equation}
 \begin{aligned} &
 {\bar \eta}^2 +p \left( \partial_{\tau}\xi^5 + \varepsilon w \partial_{x}\xi^5 - \partial_{w}\xi^4 \right)
 - \partial_{\tau}\xi^4 - \varepsilon w \partial_{x}\xi^4 = 0 \,, \\ &
 {\bar \eta}^2 +p \left( \mu \partial_{\tau}{\bar \xi}^1 - \partial_{u}{\bar \xi}^3 \right)
 + \mu \partial_{\tau}{\bar \xi}^3 + \varepsilon u p \partial_{x}{\bar \xi}^1
 + \varepsilon u \partial_{x}{\bar \xi}^3 = 0 \,, \\ &
 \varepsilon {\bar \xi}^3 - \mu \partial_{\tau}{\bar \xi}^2 + \varepsilon u \left( \mu \partial_{\tau}{\bar \xi}^1
 - \partial_{x}{\bar \xi}^2 \right)
 + \varepsilon^2 u^2 \partial_{x}{\bar \xi}^1 = 0 \,, \\ &
 \varepsilon \xi^4 - \partial_{\tau}{\bar \xi}^2 + \varepsilon w \left( \partial_{\tau}\xi^5
 - \partial_{x}{\bar \xi}^2 \right)
 + \varepsilon^2 w^2 \partial_{x} \xi^5 = 0 \,.
  \end{aligned}
 \label{detereq2}
 \end{equation}
Here, in (\ref{detereq1}) and (\ref{detereq2}) the dependencies of $\tilde \xi^i$, $\bar \xi^i$,
and $\tilde \eta^i$, $\bar \eta^i$ upon group variables are given by
 \begin{equation}
 \begin{aligned} &
 {\tilde \xi}^1 = {\tilde \xi}^1(t,\tau,x), \quad {\tilde \xi}^2 = {\tilde \xi}^2(t,\tau,x), \quad
 {\tilde \xi}^3 = {\tilde \xi}^3(t,\tau,x,u), \quad {\tilde \eta}^2 = {\tilde \eta}^2(t,\tau,x,p), \\ &
  {\bar \xi}^1 = {\bar \xi}^1(\tau,x), \quad {\bar \xi}^2 = {\bar \xi}^2(\tau,x), \quad
  {\bar \xi}^3 = {\bar \xi}^3(\tau,x,u), \quad {\bar \xi}^4 = {\bar \xi}^4(\tau,x,u),  \\ &
  {\bar \xi}^5 = \mu {\bar \xi}^1, \quad {\bar \eta}^2 = {\bar \eta}^2(\tau,x,p), \quad \eta^1=\eta^1(g), \quad
  \eta^3=\eta^3(f).
  \end{aligned}
 \label{variables1}
 \end{equation}
Second, we shall take an advantage of small parameters in
(\ref{basiceq_local}), (\ref{basiceq_nonlocal}) and, as is customary
in the approximate group analysis technique \cite{BGI_matsb_89},
express the coordinates of the group generator as power series in
$\varepsilon$ and $\mu$,
 \begin{equation}
 \xi^i = \sum\limits_{k,l=0}^{\infty}  \varepsilon^k \mu^l \xi^{i(k,l)} \,, \quad
 \eta^i = \sum\limits_{k,l=0}^{\infty}  \varepsilon^k \mu^l \eta^{i(k,l)} \,.
 \label{coord1}
 \end{equation}
Collecting terms of the same order we come to the following infinite set of equations that relate coordinates of different
orders for the fast
  \begin{equation}
 \begin{aligned}
 {\tilde \eta}^{2(k,l)} & + p \left( \partial_{t}{\tilde \xi}^{1(k,l)}
 + (1-\delta_{l,0}) \partial_{\tau}{\tilde \xi}^{1(k,l-1)}
 - \partial_{u}{\tilde \xi}^{3(k,l)} \right) \\ &
 + \partial_{t}{\tilde \xi}^{3(k,l)} +  (1-\delta_{l,0}) \partial_{\tau}{\tilde \xi}^{3(k,l-1)}
 + u p (1-\delta_{k,0}) \partial_{x}{\tilde \xi}^{1(k-1,l)} \\ &
    + u (1-\delta_{k,0})
   \partial_{x}{\tilde \xi}^{3(k-1,l)} = 0 \,, \quad k,l \geqslant 0, \\
 - \partial_{t}{\tilde \xi}^{2(k,l)} & + (1-\delta_{k,0}) {\tilde \xi}^{3(k-1,l)}
 - (1-\delta_{l,0}) \partial_{\tau}{\tilde \xi}^{2(k,l-1)} \\
 & + u (1-\delta_{k,0}) \left( \partial_{t}{\tilde \xi}^{1(k-1,l)}
 +  (1-\delta_{l,0}) \partial_{\tau}{\tilde \xi}^{1(k-1,l-1)}  - \partial_{x}{\tilde \xi}^{2(k-1,l)} \right) \\ &
 + (1-\delta_{k,0}) (1-\delta_{k,1}) u^2 \partial_{x}{\tilde \xi}^{1(k-2,l)} = 0 \,,
  \end{aligned}
 \label{detereq1a}
 \end{equation}
and the slow terms
 \begin{equation}
 \begin{aligned}
 {\bar \eta}^{2(k,l)} & + p \left( \partial_{\tau}\xi^{5(k,l)} + (1-\delta_{k,0}) w \partial_{x}\xi^{5(k-1,l)}
 - \partial_{w}\xi^{4(k,l)} \right)
  \\ & - \partial_{\tau}\xi^{4(k,l)} - (1-\delta_{k,0}) w \partial_{x}\xi^{4(k-1,l)} = 0 \,, \\
  {\bar \eta}^{2(k,l)} & + p \left( (1-\delta_{l,0}) \partial_{\tau}{\bar \xi}^{1(k,l-1)}
  - \partial_{u}{\bar \xi}^{3(k,l)} \right)
 + (1-\delta_{l,0}) \partial_{\tau}{\bar \xi}^{3(k,l-1)} \\ & + (1-\delta_{k,0}) u p \, \partial_{x}{\bar \xi}^{1(k-1,l)}
 + (1-\delta_{k,0}) u \partial_{x}{\bar \xi}^{3(k-1,l)} = 0 \,, \\
  (1-\delta_{k,0}) {\bar \xi}^{3(k-1,l)} &
  + (1-\delta_{k,0}) u \left( (1-\delta_{l,0}) \partial_{\tau}{\bar \xi}^{1(k-1,l-1)}
  - \partial_{x}{\bar \xi}^{2(k-1,l)} \right) \\ &
 - (1-\delta_{l,0}) \partial_{\tau}{\bar \xi}^{2(k,l-1)} + (1-\delta_{k,0})(1-\delta_{k,1})
 u^2 \partial_{x}{\bar \xi}^{1(k-2,l)} = 0 \,, \\
 (1-\delta_{k,0}) \xi^{4(k-1,l)} & - \partial_{\tau}{\bar \xi}^{2(k,l)}
 + (1-\delta_{k,0}) w \left( \partial_{\tau}\xi^{5(k-1,l)}
  - \partial_{x}{\bar \xi}^{2(k,l)} \right)  \\ &
 + (1-\delta_{k,0}) (1-\delta_{k,1}) w^2 \partial_{x}\xi^{5(k-2,l)} = 0 \,.
  \end{aligned}
 \label{detereq2a}
 \end{equation}
Using approximate intermediate symmetry, which follows from solutions of equations
(\ref{detereq1a})--(\ref{detereq2a}), in non-local determining equations
(\ref{detereq_gen_nonlocal}) we find a solution of the latter using variational differentiation
(see \cite{kov-de-93}, \cite[ch.4]{GrIbrKovMel_book_2010} for details) and obtain the symmetry of
the complete system (\ref{basiceq_local}), (\ref{basiceq_nonlocal}) and (\ref{basiceq2}). \par

In constructing the solution of the b.v.p. (\ref{initialcond})--(\ref{basiceq_nonlocal}) we need
not the entire set of generators but rather such a combinations of group generators that leaves
invariant the perturbation theory solution in powers of $t$ and $\tau$, the so-called
renormalization group symmetries \cite{kovsh_ufn_08}. Hence, we should specify the initial
particle distribution functions $f^0=f_{\mid t=0}$, $g^0=g_{\mid t=0}$. For concreteness, we
assume the initial velocity distribution functions to be maxwellian:
 \begin{equation}
 g^0= n^e_0(x) \exp (-u^2/2) \,, \quad f^0= n^i_0(x) \exp (-w^2/2\gamma^2)
 \,,
 \label{initialmax} \end{equation}
with the initial densities $n^e_0(x)$ and $n^i_0(x)$ and the initial zero average velocities. In
account of these initial distribution functions we have the following initial electric field
 \begin{equation}
 \begin{aligned} &
 p^0(x) = (1/\varepsilon) \int_{0}^{x} {\rm d} x \left( n^i_0(x)-n^e_0(x) \right)  \,.
  \end{aligned}
 \label{initialelfield}
 \end{equation}
Perturbation expansion of the Cauchy problem solutions in powers of $t$ and $\tau$ gives terms
$\propto O(t)$ and $\propto O(\tau)$ for the electron distribution function and $\propto O(\tau)$
for the ion distribution function, and $\propto O(t^2)$ and $\propto O(\tau^2)$ for the electric
field. Invariance conditions for these solutions specify the coordinates (\ref{coord1}) of the
group generator (\ref{gen1}). Leaving only terms that are linear in $\varepsilon$ and $\mu$ we
write these coordinates as follows
 \begin{equation}
 \begin{aligned} &
 \xi^1 = 1+ \varepsilon \tau^2 \partial_{x}\xi \,,  \quad
 \xi^2 = \varepsilon \left( \left( \delta /\Omega \right) \sin \Omega t + \mu \tau \xi \right) \,,
 \\ &
 \xi^3 =  \delta \cos \Omega t - \varepsilon \mu \tau u \partial_{x}\xi \,, \quad
 \xi^4 = \mu \left( \xi - \varepsilon \tau w \partial_{x}\xi \right) \,, \\ &
 \eta^2 =  \delta \Omega \sin \Omega t - 3 \mu \varepsilon \tau p \partial_{x}\xi  \,,
 \quad  \xi^5 = \mu \xi^1 \,, \\ &
 \eta^1 = \eta^3 = 0 \, ,
 \quad \Omega^2 = n^i(\tau,x)\equiv \int {\rm d} w f \,.
  \end{aligned}
 \label{apprsym} \end{equation}
The dependence of functions $\xi(x)$ and $\delta (x)$ upon $x$ is
expressed in terms of the initial densities distributions
$n^{e,i}_0$ and the initial electric field $p^0$,
\begin{equation}
 \begin{aligned} &
 \xi = - \varepsilon \left( (\partial_x n^e_0 / n^e_0) + \gamma^2 (\partial_x n^i_0 / n^i_0)\right)  \,,
 \quad \delta = - p^0 - \varepsilon (\partial_{x} {n^e_0} /{n^e_0}) \,, \quad
 \gamma =  V_{Ti}/c_s \,.
 \end{aligned}
  \label{xidelta} \end{equation}
For arbitrary parameters $\varepsilon$ and $\mu$ and arbitrary initial density distributions
$n^{e,i}_0$ formulas (\ref{apprsym}) describe the approximate symmetry. However, in two limiting
cases infinite series (\ref{coord1}) terminate and we get the exact symmetry group. The first
case is referred  to  electron plasma with neutralizing homogeneous ion background ($\mu = 0$,
$\Omega^2 = n^i_0 = const$) \cite{Tar_jtf_76,GrigMel_dan_87,GrigMel_jma_95}, which gives the
generator
  \begin{equation}
  X = \partial_{t} +
     \varepsilon \left( \delta /\Omega \right) \sin \Omega t \partial_{x} +
     \delta \cos \Omega t \partial_{u} +
     \delta \Omega \sin \Omega t \partial_p \,.
  \label{genelplasma} \end{equation}
The second case is referred to quasi-neutral approximation for electron-ion plasma with zero
current and charge densities $j=\rho=0$ \cite{BychKovTikh_jetf_02} that is realized for  $\delta
=0$ and the initial gaussian densities distribution, $\xi \varpropto \beta x$
  \begin{equation}
  \begin{aligned} &
  X = \left( 1+ \beta \varepsilon \tau^2 \right) \partial_{\tau}  +
     \varepsilon \beta \tau x \partial_{x}  +
      \beta \left(  {\mu x} - \varepsilon \tau u \right) \partial_{u}  +
     \beta \left( x - \varepsilon \tau w \right) \partial_{w}
     - 3 \varepsilon \beta \tau p \partial_p \,.
    \end{aligned}
 \label{quasineutral} \end{equation}
The additional term in $\xi^3$ in (\ref{quasineutral}) that refers to acceleration of electrons
is omitted in (\ref{apprsym}) as it is of the higher order $O(\mu^2)$ as compared to that
included in (\ref{apprsym}).

\section{Lie equations \label{sec4}}
To construct group invariant solution for the b.v.p.
(\ref{initialcond})--(\ref{basiceq_nonlocal}) we should find solutions of the Lie equations for
the group generator (\ref{gen1}) with coordinates (\ref{coord1}),
  \begin{equation}
  \begin{aligned} &
 \frac{d t}{d a} = 1 + \varepsilon \tau^2 \partial_{x}\xi \,, \quad t_{\mid
 a=0}=t^{\prime}\,, \\ &
 \frac{d x}{d a} = \varepsilon \left( \left( \delta /\Omega \right) \sin \Omega t + \mu \tau \xi \right) \,,
 \quad x_{\mid
 a=0}=x^{\prime}\,, \\ &
  \frac{d u}{d a} = \delta \cos \Omega t - \varepsilon \mu \tau u \partial_{x}\xi \,, \quad u_{\mid
 a=0}=u^{\prime}\,, \\ &
  \frac{d w}{d a} = \mu \left( \xi - \varepsilon \tau w \partial_{x}\xi \right) \,, \quad w_{\mid
 a=0}=w^{\prime}\,, \\ &
   \frac{d \tau}{d a} = \mu \left( 1 + \varepsilon \tau^2 \partial_{x}\xi  \right) \,, \quad \tau_{\mid
 a=0}=\tau^{\prime}\,, \\ &
 \frac{d p}{d a} = \delta \Omega \sin \Omega t - 3 \mu \varepsilon \tau p \partial_{x}\xi \,, \quad p_{\mid
 a=0}=p^{\prime}\,, \\ &
 \frac{d f}{d a} = \frac{d g}{d a} = 0\,, \quad g_{\mid
 a=0}=g^{\prime}\,, \quad f_{\mid
 a=0}=f^{\prime}\,.
   \end{aligned}
  \label{lieeqgen} \end{equation}
Solution of the b.v.p. (\ref{initialcond})--(\ref{basiceq_nonlocal}) are expressed as usual in
terms of invariants of the group (\ref{gen1}), (\ref{coord1}) that result from solutions of
(\ref{lieeqgen}) after excluding the group parameter $a$. Due to a difference in characteristic
time scales we can separate ``fast'' and ``slow'' group invariants, applying the averaging
procedure to Lie equations. \par

In fact, at small time $ t > 0 $, $ 1/ \mu \gg t \gg 1/\Omega$, the ''ion`` terms that are
$\propto \mu$ can be omitted and we come to simplified Lie equations (equations for group
invariants $f$, $g$, $\tau$, $w$ are omitted here),
  \begin{equation}
  \begin{aligned} &
 \frac{d t}{d a} = 1 \,, \quad t_{\mid
 a=0}=t^{\prime}\,; \quad  \frac{d x}{d a} = \varepsilon \left( \delta /\Omega \right) \sin \Omega t  \,,
 \quad x_{\mid
 a=0}=x^{\prime}\,; \\ &
  \frac{d u}{d a} = \delta \cos \Omega t \,, \quad u_{\mid
 a=0}=u^{\prime}\,; \quad \frac{d p}{d a} = \delta \Omega \sin \Omega t \,, \quad p_{\mid
 a=0}=p^{\prime}\,,
   \end{aligned}
  \label{lieeqfast} \end{equation}
the solutions of which define invariants of ``fast'' motions at small time $t \ll 1/\mu$:
 \begin{equation}
 \begin{aligned} &
  J_1 = p+\delta \cos \Omega t \equiv - \varepsilon (\partial_x n^e_0 / n^e_0)|_{x = x^\prime}
  \,, \\ &
  J_2 = x + (\varepsilon \delta/\Omega^2) \cos \Omega t \equiv
  x^\prime + (\varepsilon \delta (x^\prime)/\Omega^2(x^\prime)) \,, \\ &
  J_3 = u -  (\delta/\Omega ) \sin \Omega t \equiv u^\prime  \,, \\ &
  J_4 = g \equiv g^0(x^\prime,u^\prime) \,.
 \end{aligned}
 \label{fastinv} \end{equation}
On the contrary, averaging the complete Lie equations on a large
time scale $T \gg 1/\mu $ we come to Lie equations defining ``slow''
motions (equations for group invariants are again omitted),
  \begin{equation}
  \begin{aligned} &
 \frac{d \tau}{d a} = 1 + \varepsilon \tau^2 \partial_{\bar x}{\bar \xi} \,, \quad
 \frac{d {\bar x}}{d a} = \varepsilon \tau {\bar \xi} \,, \quad \frac{d {\bar p}}{d a} =
   - 3 \varepsilon \tau {\bar p} \partial_{\bar x}{\bar \xi} \,,  \\ &
  \frac{d {\bar u}}{d a} =  - \varepsilon \tau {\bar u} \partial_{\bar x}{\bar \xi} \,, \quad
  \frac{d w}{d a} = {\bar \xi} - \varepsilon \tau w \partial_{\bar x}{\bar \xi} \,,
   \end{aligned}
  \label{lieeqslow} \end{equation}
with the corresponding ``slow'' invariants
 \begin{equation}
 \begin{aligned} &
  I_1 = \bar{f} \equiv {f}^{\, \prime} \,, \quad
  I_2 = \bar{g} \equiv  {g}^{\, \prime} \,, \quad
  I_3 = \bar{p}{\bar \xi}^{3} \equiv {p}^{\, \prime} {{\xi}^{\prime}}^{3}  \,,  \\ &
  I_4 = \frac{\varepsilon \tau^2}{2 {\bar \xi}^2}  -  \int^{\bar x}{\rm d} z/ {\xi}^{\,3} \equiv
  \frac{\varepsilon {\tau^{\prime}}^2}{2 {\xi^{\prime}}^2}  -  \int^{x^{\prime}}{\rm d} z/ {\xi}^{\,3}
  \,,  \quad
  I_5 = {\bar \xi} \bar{u} \equiv  {\xi}^{\prime} {u}^{\, \prime} \,, \\ &
  I_6 = {\bar \xi} \bar{w} - \frac{1}{\sqrt{2 \varepsilon }} \int^{\bar x} {\rm d} y
  \left( \int^{y}{\rm d} z / {\xi}^{\, 3} \right)^{-1/2} \equiv {\xi}^{\prime} {w}^{\, \prime}
  - \frac{1}{\sqrt{2 \varepsilon }} \int^{s} {\rm d} y
  \left( \int^{y}{\rm d} z / {\xi}^{\, 3} \right)^{-1/2} \,.
  \end{aligned}
  \label{slowinv} \end{equation}
Here the ``primed'' variables are related to values at $\tau \to 0$
 \begin{equation}
 \begin{aligned} &
  {f}^{\, \prime} = f^0(s,w^{\, \prime}) \,, \
  {g}^{\, \prime} = g^0(s,u^{\, \prime}) \,, \
  {p}^{\, \prime} = - \varepsilon \partial_{s}\varphi \,, \
  s= x^{\prime} + \frac{\varepsilon \delta(x^{\prime})}{ \Omega^2(x^{\prime})} \,, \\ &
 \varepsilon^2 \partial_{xx}\varphi +  n^i_0(x) - {\rm e}^{\varphi} = 0 \,, \ \partial_{x}\varphi|_{x = 0} =
 \partial_{x}\varphi|_{x \to \infty } = 0\,, \ \varphi|_{x = 0} = C < \infty.
  \end{aligned}
  \label{variablesprime} \end{equation}
In the next section we use the fast and slow invariants to construct
analytical solutions of the Cauchy problem for the kinetic equations
(\ref{initialcond})--(\ref{basiceq_nonlocal}).

\section{Slow dynamics of plasma particles \label{sec5}}

Let we consider the slow dynamics of plasma particles under simplifying assumptions, small value
of $p_0 < 1 $,  and low ion temperature, $\gamma \to 0$. Then, following
(\ref{slowinv})--(\ref{variablesprime}), $s \thickapprox  x^\prime$ and $\xi$ coincides with
$p^\prime = {\bar p}|_{\tau = 0}$, and we come to simplified expressions, which define dynamics
of plasma ions,
  \begin{equation}
  \begin{aligned} &
  \bar{f} = f^0(x^{\, \prime},w^{\, \prime})  \,, \quad
   \bar{g} = g^0(x^{\, \prime},w^{\, \prime})  \,, \quad
  \bar{p} =  \left( {\xi}^{\prime}/{\bar \xi} \right)^{3} {p}^{\, \prime}(x^{\, \prime}) \,, \quad
  \varepsilon \tau^2  = {2 {\bar \xi}^2}  \int_{x^{\, \prime}}^{\bar x}{\rm d} z/ {\xi}^{\,3} \,,
  \\ &
    \bar{w} =  \left( {\xi}^{\prime}/{\bar \xi} \right) {w}^{\, \prime}
  + \frac{1}{\bar \xi \sqrt{2 \varepsilon }} \int_{x^{\, \prime}}^{\bar x} {\rm d} y
  \left( \int_{x^{\, \prime}}^{y}{\rm d} z / {\xi}^{\, 3} \right)^{-1/2}
    \,, \quad  \bar{u} =  \left( {\xi}^{\prime}/{\bar \xi} \right) {u}^{\, \prime}.
 \end{aligned}
 \label{finiteslow} \end{equation}
For completeness we also present global characteristics for plasma
ions, their average velocity $v^i_{av}$, density $n^i_{av}$ and
temperature $T^i$,
  \begin{equation}
 \begin{aligned} &
  v^i_{av} = \frac{1}{\bar \xi \sqrt{2 \varepsilon }} \int_{x^{\prime}}^{\bar x} {\rm d} y
  \left( \int_{x^{\prime}}^{y}\frac{{\rm d} z}{ {\xi}^{\, 3}} \right)^{-1/2}, \
   n^i_{av} = {n}^{i}_{0}(x^{\prime}) \frac{ {\xi}^{\prime}}{{\bar \xi} } \,, \
  T^i = T^{i}_{0}\left( \frac{{\xi}^{\prime}}{{\bar \xi}} \right)^{2}\,.
 \end{aligned}
 \label{iondensaver} \end{equation}
These formulas are analyzed below for two distinct initial electric field and density distributions.

\subsection{ Examples of slow plasma dynamics: gaussian density profile}
\begin{figure}[h!]
 \includegraphics[width=0.98\textwidth]{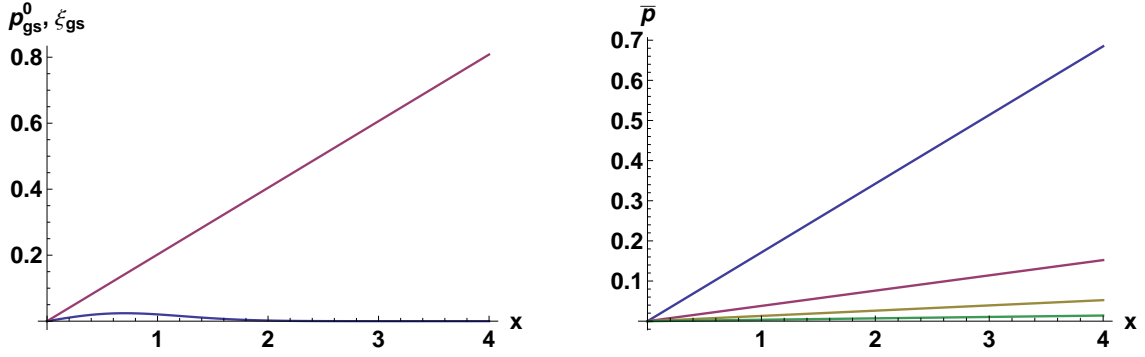}
 \caption{Distribution of electric field $p^0_{gs}$ (blue line) and $\xi_{gs}$ (red line)
 for $\tau=$0 (left), and ``slow'' electric field $\bar p$ (right) for $\tau=2$(blue line),
 $\tau=8$(red line), $\tau=12$(yellow line), $\tau=18$(green line), and for $b=1.01$,
 $\varepsilon = 0.1$, $\mu=\sqrt{1/2000}$ and
 $\gamma =0.1$.}
  \label{p0gs}
 \end{figure}

We start with a specific situation when electron and ion densities balance each other, and are
described by gaussian curves,
 \begin{equation}
 \begin{aligned} &
  {n}_e^0(x)= n_i^0(x) = (1/\sqrt{\pi}) \exp (- x^2) \,, \quad
  \xi= 2 \varepsilon x \left( 1 + \gamma^2 \right) \,.
 \end{aligned}
 \label{xigauss} \end{equation}
Substituting expressions (\ref{xigauss}) in (\ref{finiteslow})--(\ref{iondensaver}) we get formulas for finite group
transformations for ``slow'' variables
 \begin{equation}
 \begin{aligned} &
   \bar{p} =  \frac{{p}^{\, \prime}}{(1+ \nu^2 \tau^2 )^{3/2}}   \,,
  \quad
   \bar{x} = x^{\prime} \sqrt{1+ \nu^2 \tau^2 } \,, \quad  \nu^2 = 2 \varepsilon^2 \left( 1 + \gamma^2
  \right) \,,
  \\ &
   \bar{u} =  \frac{{u}^{\prime} }{\sqrt{1+ \nu^2 \tau^2 }}  \,,
   \quad
  \bar{w} =  \frac{1}{\sqrt{1+ \nu^2 \tau^2 }} \left( {w}^{\prime}
             + \frac{\nu^2}{\varepsilon} {x}^{\, \prime} \tau \right)  \,,
             \quad  \bar{f} =  {f}^{\, \prime}  \,, \quad
  \bar{g} =  {g}^{\, \prime}
   \,.
  \end{aligned}
 \label{slowfinite} \end{equation}
and formulas for ion average density, velocity and temperature
 \begin{equation}
 \begin{aligned}
 & n^i_{av} =   \frac{1}{\sqrt{\pi(1+ \nu^2 \tau^2)}} \exp \left( -
 \frac{{\bar x}^2}{1+ \nu^2 \tau^2} \right) \,, \\ &
  v^i_{av} = \frac{ {\bar x}}{\varepsilon }  \frac{\nu^2  \tau }{(1+ \nu^2 \tau^2)}\,, \
  T^i = \frac{T^{i}_{0}}{ \left( 1+ \nu^2 \tau^2 \right)^{2}} \,.
 \end{aligned}
 \label{slowaverage} \end{equation}
These formulas demonstrate the self-similar dependence of the ion density -- the Gaussian-type
density distribution is preserved though the spatial scale of this distribution as well as the
maximum of the ion density varies. The spatial dependence of the average ion velocity is linear,
and the ion temperature is uniform in space and monotonically decreases with the growth of time
$\tau$, as it was demonstrated in~\cite{BychKovTikh_jetf_02}. In case, when the initial ion
density varies from the electron density, i.e. for $ n_i^0(x)= (b/\sqrt{\pi}) e^{-b^2 x^2}$, $b
\thickapprox 1$, the above formulas are still valid, provided $1+\gamma^2$ is replaced by $1 +
b^2\gamma^2$. The difference between the initial densities distributions of particles leads to
nonzero values of $p^0$,
 \begin{equation}
 \begin{aligned} &
  p^0_{gs} = \frac{1}{2 \varepsilon} \left( {\rm Erf}(bx) - {\rm Erf}(x)  \right),
  \
  \xi_{gs}= 2 \varepsilon x \left( 1 + b^2 \gamma^2 \right) \,, \ \delta^0 = -p^0_{gs} + 2 \varepsilon
  x \,.
 \end{aligned}
 \label{pxigs} \end{equation}
however for $b$ close to unity the values of $p^0$ are small as compared to $\xi$.
Figure~\ref{p0gs} (left panel) illustrates this fact, while the right panel shows the spatial
distribution of the ``slow'' electric field $\bar p$ at different nonzero time moments.

\begin{figure}[h!]
 \includegraphics[width=0.98\textwidth]{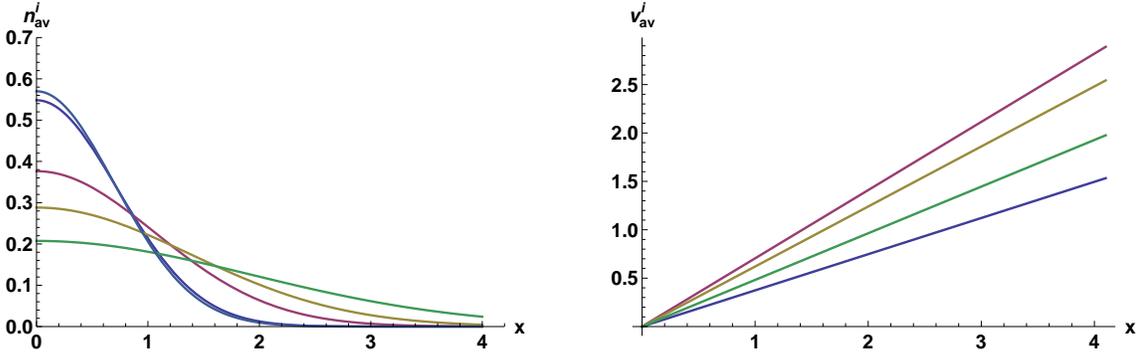}
  \caption{Averaged density and velocity distributions of ions for $\tau=0$ (light-blue line),
  $\tau=2$(dark-blue line), $\tau=8$(red line), $\tau=12$(yellow line), $\tau=18$(green line), and
  $b=1.01$, $\varepsilon = 0.1$, $\mu=\sqrt{1/2000}$, and $\gamma =0.1$.} \label{gaussfig}
  \end{figure}

\noindent As for the spatial dependencies of the average ion velocity and density at the same
time moments they are presented on Figure~\ref{gaussfig}.
It follows from (\ref{pxigs}) that the oscillating electric field is of the order of the average
electric field that accelerates ions. In the next section we consider the opposite situation when
the initial electric field practically concise with the initial ``slow'' electric field $\bar p$
and the amplitude of the ``fast'' electric field is small.

 \subsection{Examples of slow plasma dynamics: Lorentz density
         profile}

Turn now to the case when there is very slight difference between $p^0$ and $\xi$ that is realized, for example, for the
Lorentz-type initial density distribution,
 \begin{equation}
 \begin{aligned} &
  n_e^0(x)= (1/ \pi (1 + x^2)) \,, \quad
  n_i^0(x)= (b/\pi ( 1+  b^2 x^2)) \,, \quad \mid  b-1 \mid \, \ll 1 \,.
 \end{aligned}
 \label{lorentz1} \end{equation}
Substitution of (\ref{lorentz1}) into (\ref{initialelfield}) gives the following formulas for the
spatial distribution of the initial electric field and the function $\xi$,
 \begin{equation}
  \begin{aligned} &
  p^0_l(x)= \frac{1}{\pi \varepsilon} \left( \arctan bx - \arctan x \right) \,, \quad
  \xi_l= 2 \varepsilon x \left( \frac{1}{1 + x^2} + \frac{\gamma^2 b^2}{1 + b^2 x^2}  \right) \,.
 \end{aligned}
 \label{lorentzxi} \end{equation}
 \begin{figure}[h!]
 \includegraphics[width=0.98\textwidth]{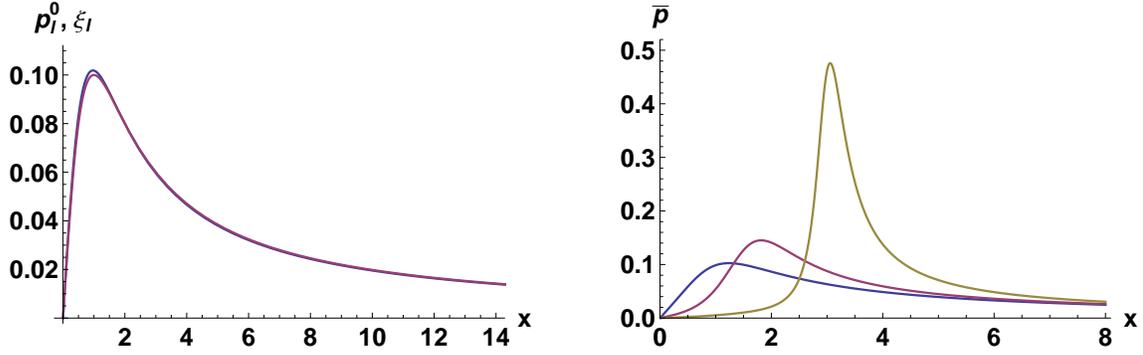}
  \caption{Plots of $p^0_l$ (blue line) and $\xi_l$ (red line) at $\tau=$ 0 (left),
 and ``slow'' electric field $\bar p$ (right) at $\tau=$~4 (blue line), $\tau=$ 10 (yellow line),
 $\tau=$ 18 (red line), for $a=1.0$, $b=1.0661$, $\varepsilon = 0.1$, $\mu=\sqrt{1/2000}$ and
 $\gamma =0.001$.}
 \label{lorp0}
  \end{figure}
The left panel of Figure~\ref{lorp0} demonstrates the difference between $p^0_l$ and $\xi_l$, and
the right panel shows the spatial distribution of the ``slow'' electric field $\bar p$ at
different moments of time $\tau$.

\begin{figure}[hbt!]
 \includegraphics[width=1.02\textwidth]{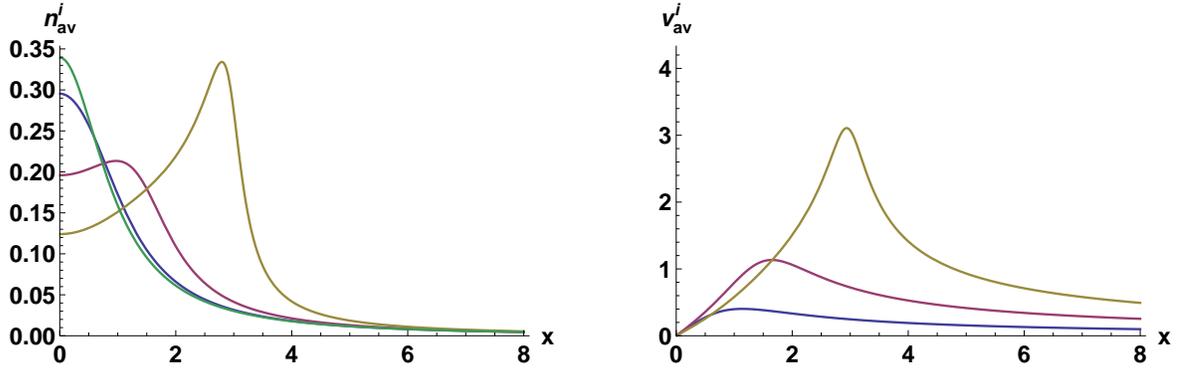}
 \caption{Density and velocity distributions for ions at $\tau=$ 0 (green line), $\tau=$~4 (blue line),
 $\tau=$ 10 (red line), $\tau=$ 18 (yellow line), for $b=1.0661$, $\varepsilon = 0.1$, $\mu=\sqrt{1/2000}$ and
 $\gamma =0.001$.}
 \label{denvellor}
 \end{figure}

As for the average ion density, temperature  and velocity they are given by the formulas
 \begin{equation}
  \begin{aligned} &
  n^i_{av} =  \frac{b}{\pi ( 1+  b^2 {x^{\prime}}^2)} \frac{{\xi}^{\prime}_l}{{\bar \xi_l} }
  \,, \
  T^i = T^{i}_{0} \left( \frac{{\xi}^{\prime}_l}{{\bar \xi_l} } \right)^2
  \,, \
  v^i_{av} = \frac{1}{\bar \xi_l \sqrt{2 \varepsilon }} \int_{x^{\prime}}^{\bar x} {\rm d} y
  \left( \int_{x^{\prime}}^{y}{\rm d} z / {\xi}^{\, 3}_l \right)^{-1/2}
   \,.
 \end{aligned}
 \label{veldenslor} \end{equation}
and are plotted on the Figure~\ref{denvellor}.

\section{Fast dynamics of particles \label{sec6}}

\begin{figure}[h!]
 \includegraphics[width=0.98 \textwidth]{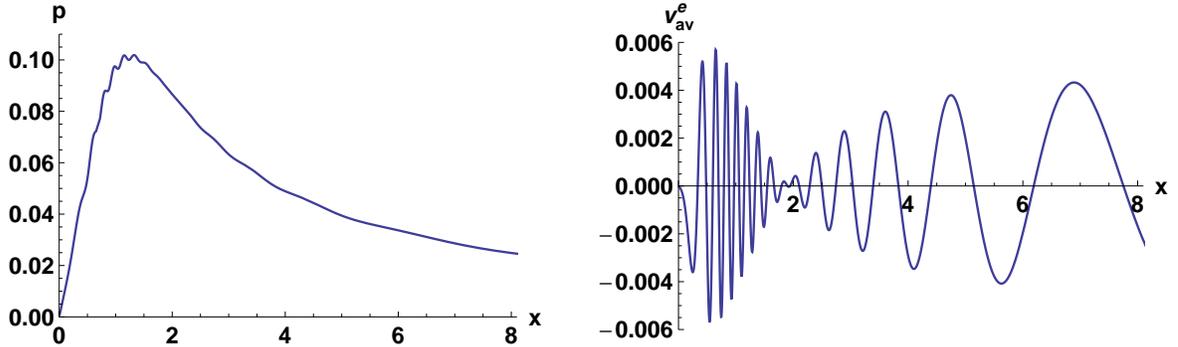}
 \caption{Electric field and average electron velocity distributions at $\tau=4$ for $a=1.0$, $b=1.0661$,
 $\varepsilon = 0.1$, $\mu=\sqrt{1/2000}$ and $\gamma =0.001$.}
 \label{velfield_lor_16}
 \end{figure}

In this section we use slow invariants to restore the complete dynamics of fast particles,
electrons. For clarity's sake we consider the case of small values of $ \delta \varepsilon \ll
1$, which means that $x$ is identical to $\bar x$, and rewrite the Lie equations (\ref{lieeqgen})
in a simplified form
 \begin{equation}
  \begin{aligned} &
 \frac{d t}{d a} = 1 + \varepsilon \tau^2 \partial_{x}\xi \,, \quad
 \frac{d (\xi u) }{d a} = \delta \xi \cos \Omega t \,, \quad
 \frac{d (\xi^3 p) }{d a} = \delta \Omega \xi^3 \cos \Omega t \,.
   \end{aligned}
  \label{lieeqfast1} \end{equation}
According to the procedure of averaging \cite{KBM_bk_74} we can write the solutions of equations
(\ref{lieeqfast1}) by integrating over fast time $t$ and taking into account the dependence upon
slow time by including the dependence upon $\bar x$ and  $\tau$  into $\delta$ and $\xi$.
However, the enhanced precision is achieved by direct integration of the Lie equations
(\ref{lieeqfast1}) in account of the slow dependence of $\tau$ upon $\bar x$ as given by slow
motion invariants
\begin{equation}
 \begin{aligned} &
  p =  \frac{{{\xi}^{\prime}}^3}{{\bar \xi}^3} {\bar p}^{\, \prime}
  + \frac{1}{{\bar \xi}^3} \int\limits_{x^\prime}^{x} {\rm d} x^{\prime\prime}
  \frac{\bar \delta {\bar\Omega}{\xi}^2(x^{\prime\prime})}{\varepsilon \mu \tau(x^{\prime\prime})}
     \sin ( {\bar\Omega} t(x^{\prime\prime})) \,, \\ &
  u =  \frac{{\xi}^{\prime}}{{\bar \xi}} {\bar u}^{\, \prime}
     +  \frac{1}{{\bar \xi}}\int\limits_{x^\prime}^{x} {\rm d} x^{\prime\prime}
     \frac{\bar \delta }{\varepsilon \mu \tau(x^{\prime\prime})}
     \cos ( {\bar\Omega} t(x^{\prime\prime})) \,.
  \end{aligned}
 \label{fast2} \end{equation}
Electron distribution function $g=g^{\prime}\equiv g^0(x^{\prime},u^{\prime})$ is the invariant
of group transformations. Thus substituting $x={\bar x}$ and $u^{\prime}$ from (\ref{fast2}) in
$g^0$ and integrating over the velocity $u$ gives the integral characteristic, the average
electron velocity and density
\begin{equation}
 \begin{aligned} &
 n^e_{av} = {n}^{e}_{0}(x^{\prime}) \left( {\xi}^{\prime}/{\bar \xi} \right)
             \,, \quad
  u^e_{av} =  \frac{1}{\xi^{\prime}}\int\limits_{x^\prime}^{x} {\rm d} x^{\prime\prime}
  \frac{\bar \delta }{\varepsilon \mu \tau(x^{\prime\prime})}
     \cos ( {\bar\Omega} t(x^{\prime\prime})) \,.
  \end{aligned}
 \label{neveaver} \end{equation}
To illustrate these formulas we employ results of the previous section and consider the
Lorentz-type initial densities profiles (\ref{lorentz1}) with the function $\xi=\xi_l$ defined by
(\ref{lorentzxi}). Substituting $\xi_l$ in (\ref{fast2})--(\ref{neveaver}) we obtain the formulas
for the spatial distribution of the electric field and the average electron velocity that are
presented on the figures below for three different time moments. Figure~\ref{velfield_lor_16}
corresponds to moderate values of $\tau=4$, when the ion density is concentrated mainly in the
center of the bunch thus leading to small-scale spatial oscillations primarily in this region.
As the bunch spreads with growth of $\tau$ the small-scale spatial oscillations moves outward as shown on
Figure~\ref{velfield_lor_100} for $\tau=10$,
 {\begin{figure}[h!]
 \includegraphics[width=0.98\textwidth]{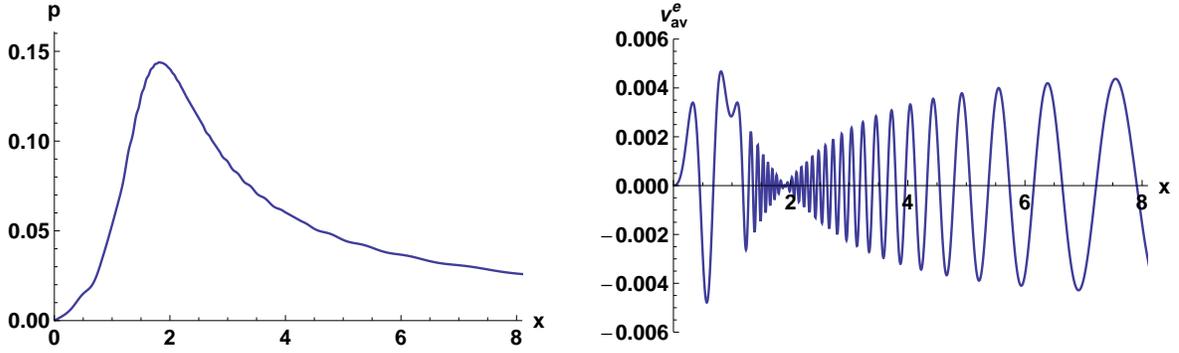}
 \caption{Electric field and average electron velocity distributions at $\tau=10$ for $a=1.0$, $b=1.0661$,
  $\varepsilon = 0.1$, $\mu=\sqrt{1/2000}$ and $\gamma =0.001$.}
 \label{velfield_lor_100}
  \end{figure}
and on Figure~\ref{velfield_lor_324} for $\tau=18$. The figures also show that the ``complete''
electric field $p$ in this case oscillates with the same spatial period as the mean electron
velocity and only slightly differs from the average electric field $\bar p$ (compare with
Figures~\ref{lorp0}).
 {\begin{figure}[t!]
 \includegraphics[width=0.98\textwidth]{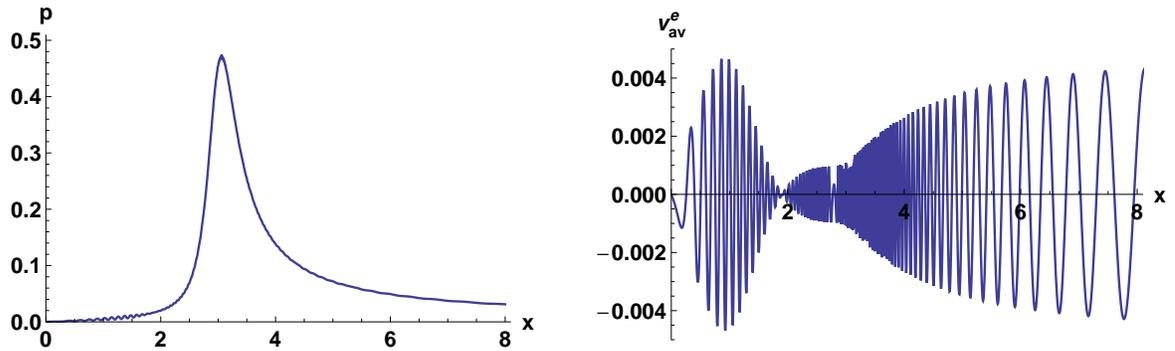}
 \caption{Electric field and average electron velocity distributions at $\tau=18$ for $a=1.0$, $b=1.0661$,
 $\varepsilon = 0.1$, $\mu=\sqrt{1/2000}$ and  $\gamma =0.001$.} \label{velfield_lor_324}
 \end{figure}

 \section{Conclusion}
In the above analysis of a particular physical problem, expansion of a plasma bunch, a new
promising tool for analyzing nonlinear multi-scale systems was considered. The main idea consists
in employing the Krylov-Bogoliubov-Mitropolskii procedure of averaging to construct solutions of
Lie equations. The procedure of separating ``fast'' and ``slow'' terms in coordinates of group
generator naturally occurs in determining equations while constructing the symmetry for nonlinear
equations that describe multi-scale behavior of any physical system. In our consideration we use
the averaging procedure in combination with a perturbation technique of group analysis
\cite{BGI_matsb_89} that gives approximate symmetries for the analyzed problem and helps to
construct approximate RG-invariant solution for arbitrary initial distribution functions of
particles.
\par

The use of the averaging procedure in modern group analysis naturally separates
invariant manifolds, related to slow and fast Lie equations into slow and fast invariant
manifolds. This separation is in the root of the theorem of invariant representation
\cite[\S18]{Ovs_bk_78}: averaging the fast invariant solution that appears as an
oscillating ``curve'' on fast manifold yields a smooth curve on the slow invariant
manifold as shown in the previous section (compare to the method of slow invariant
manifold for  describing kinetics of dissipative systems \cite{Gorban_book_2005}). The
merits of the approach with different scales that simplifies both the procedure of
finding the admitted group and construction of the group invariant solutions point to
the quest for future potential applications.

\section*{Acknowledgments}

This research was fulfilled during the visit to Ufa State Aviation Technical University in
December 2011. The author acknowledges a financial support from the University of the research in
the framework of the mega-grant No.~11.G34.31.0042. The author would like to express his
gratitude to Prof. V.Yu.~Bychenkov and Prof. D.V.~Shirkov for many stimulating conversation. This
research has been partially supported by the presidential grant Scientific School No.~3810.2010.2
also.


\end{document}